\documentclass[graybox]{svmult}

\usepackage{type1cm}        % activate if the above 3 fonts are
\usepackage{makeidx}         % allows index generation
\usepackage{graphicx}        % standard LaTeX graphics tool
\usepackage{multicol}        % used for the two-column index
\usepackage[bottom]{footmisc}% places footnotes at page bottom

\usepackage{newtxtext}       % 
\usepackage{newtxmath}       % selects Times Roman as basic font

\usepackage{enumerate, commath}
\usepackage{natbib}

\newcommand{\cG}{\mathcal{G}}

\newcommand{\cS}{\mathcal{S}}

\usepackage{accents}
\newcommand{\ubar}[1]{\underaccent{\bar}{#1}}

\makeindex             % used for the subject index
\begin{document}

\title*{Distribution-Free Pointwise Adjusted $P$-Values for Functional Hypotheses}
\author{Meng Xu and Philip T.~Reiss}
\institute{Meng Xu \at Department of Statistics, University of Haifa, Haifa 31905, Israel \email{mxu@campus.haifa.ac.il}
\and Philip T.\ Reiss \at Department of Statistics, University of Haifa, Haifa 31905, Israel \email{reiss@stat.haifa.ac.il}}
\index{Xu, M.} %First Author
\index{Reiss, P.~T.} %Second Author

\maketitle

\abstract{Graphical tests assess whether a function of interest departs from an envelope of functions generated under a simulated null distribution. This approach originated in spatial statistics, but has recently gained some popularity in functional data analysis. Whereas such envelope tests examine deviation from a functional null distribution in an omnibus sense, in some applications we wish to do more: to obtain $p$-values at each point in the function domain, adjusted to control the family-wise error rate. Here we derive pointwise adjusted $p$-values based on envelope tests, and relate these to previous approaches for functional data under distributional assumptions. We then present two alternative distribution-free $p$-value adjustments that offer greater power. The methods are illustrated with an analysis of age-varying sex effects on cortical thickness in the human brain.
}

\section{Introduction}
In many functional data analysis (FDA) settings, one wishes to test either a null hypothesis
\begin{equation}\label{onehyp}H_0: f(s)=0\mbox{ for all }s\in\cS,\end{equation}
for a function $f$ defined on a domain $\cS$, or alternatively a family of null hypotheses \begin{equation}\label{h0s}\{H_0(s): s\in\cS\}\end{equation} 
where for each $s$, $H_0(s)$ is the pointwise hypothesis $f(s)=0$. For example, $f$ may refer to
\begin{enumerate}[(i)]
\item a group difference $f(s)=g_1(s)-g_2(s)$, where $g_1,g_2$ denote mean functions in two subsets of a population, or
\item a coefficient function $f(s)=\beta(s)$ in a functional linear model.
\end{enumerate}

Clearly the global hypothesis $H_0$ in \eqref{onehyp} is just the intersection over all $s$ of the pointwise hypotheses $H_0(s)$ in \eqref{h0s}. The difference is that whereas \eqref{onehyp} refers to a single test, for which a single $p$-value would be appropriate, the family \eqref{h0s} gives rise to a collection of $p$-values. The latter setup is appropriate when the values of $f(s)$ for different $s$ carry distinct scientific meaning. For example,  in \secref{appsec} below we test for sex-related differences in the thickness of the human cerebral cortex as a function of age $s$. In this context, age-specific results may have implications for the study of brain development.  

Previous work has tended to focus either on distribution-free tests of the global hypothesis \eqref{onehyp} (see \secref{envsec} below), or on multiplicity-adjusted parametric pointwise tests for the family \eqref{h0s}. 
As we show in \secref{adjsec}, it is straightforward to combine the advantages of both approaches---that is, to derive pointwise adjusted $p$-values without having to specify a null statistic distribution. In \secref{morepower}, we present two alternative pointwise $p$-value adjustments that offer improved power. 

\section{Setup}
We let $T(s)$ ($s\in\cS$) denote a functional test statistic for null hypothesis \eqref{onehyp}, and take as given a group of permutations of the data, along with the null hypothesis that the joint distribution of $T(s)$, $s\in\cS$, is invariant to such permutations. This hypothesis may be stronger than \eqref{onehyp}, but for the sake of a brief and general presentation, we ignore that distinction here. 
Let $T_0$ be the test statistic function computed with the real data,  and $T_1,\ldots,T_{M-1}$ be test statistic functions that are computed with randomly permuted data sets and thus constitute a simulated null distribution. 
We consider $T_0(s),\ldots,T_{M-1}(s)$ only for $s\in\cG$, for a finite set $\cG\subset\cS$  (e.g., a grid of points spanning $\cS$, if the latter is a subinterval of the real line). We assume $\cG$ to be an adequate approximation to $\cS$, in the sense that the difference between a minimum over $\cG$ versus over $\cS$ is negligible \citep[see][for a relevant treatment of grid approximations in functional hypothesis testing]{cox2008}.  We further assume that there are no pointwise ties, i.e., ties among $T_0(s),\ldots,T_{M-1}(s)$ for a given $s\in\cG$.%\footnote{We need to be clearer about the role of $\cG$ throughout the paper---can we just assume that minimum over $\cG$ equals minimum over $\cS$?}

\section{Envelope tests}\label{envsec}
Hypotheses regarding spatial point patterns are commonly tested by functions $T(s)$ of interpoint distance $s$, such as the $K$ function of \cite{ripley1977}. Such functions typically have unknown null distributions, and hence are most readily tested via Monte Carlo methods. This is the motivation for graphical or envelope tests \citep{ripley1977,davison1997,baddeley2014}, which have recently been formalized, extended, and applied to functional data \citep{myllymaki2017,mrkvicka2018}.   

The global envelope test (GET) of \cite{myllymaki2017}  is based on the ranks $R_m^*(s)$ of $T_m(s)$ among $T_0(s),\ldots,T_{M-1}(s)$ for $s\in\cG$. Here rank is defined in such a way that low rank indicates maximal inconsistency with the null hypothesis. %for instance, for a one-sided test with larger $T(s)$ meaning greater departure from the null, rank 1 means largest.
Thus, depending on the test, $R_m^*(s)$ may be rank be from smallest to largest, rank from largest to smallest, or for a two-sided test, the smaller of the two. The minimum rank attained by $T_m$, $R_m=\min_{s\in\cG}R_m^*(s)$,  is a functional depth \citep{lopez2009}, which we may call the min-rank depth.
  The GET $p$-value is then defined as
\begin{equation}\label{pp}p_+=\frac{\sum_{m=1}^{M-1} \mathbb{I}(R_m \leq R_0)  +1}{M}.\end{equation}
This $p$-value has a graphical interpretation in terms of envelopes, which we define here in a manner that is consistent with  \cite{myllymaki2017}, but that relates to $p$-values rather than a specified level $\alpha$. For $j\geq 1$, let $\kappa_j=\sum_{m=0}^{M-1}\mathbb{I}(R_m\leq j)$, and let $E^{\kappa_j}$ be the envelope defined by the set of  $M-\kappa_j$ curves $\{T_m: R_m > j\}$, %m\in I_{\kappa_j}\}$, where $I_{\kappa_j}=\{m:R_m > j\}$,
  that is, the range from $\ubar{T}^{\kappa_j}(s)=\min_{m: R_m > j}T_m(s)$ to $\bar{T}^{\kappa_j}(s)=\max_{m: R_m > j}T_m(s)$ for each $s$. We say that $T_0$ exits this envelope at $s$ if $T_0(s)\notin[\ubar{T}^{\kappa_j}(s),\bar{T}^{\kappa_j}(s)]$. Arguing as in \cite{myllymaki2017}, one can show that $p_+\leq \kappa_j/M$ if and only if $T_0$ exits $E^{\kappa_j}$ at some $s$.

\section{Adjusted $p$-values}\label{adjsec}
Turning from the single hypothesis \eqref{onehyp} to the family \eqref{h0s} of pointwise hypotheses, the na\"ive or  raw permutation-based $p$-values are 
\begin{equation}\label{rawp}p(s)=R_0^*(s)/M\end{equation} 
for each $s$. These $p$-values, however, require adjustment for multiplicity \citep{wright1992} in order to control the overall type-I error rate, usually taken as the family-wise error rate (FWER). %While a rigorous general definition of adjusted $p$-values is elusive \citep{yekutieli1999},\footnote{Instead of this blablabla, say something intelligent about FWER control and the meaning of $p$-value adjustment.} 
 Strictly speaking, since the GET is a single test as opposed to a multiple testing procedure,  adjusted $p$-values with respect to the GET are undefined. But it is natural to define the GET-adjusted $p$-value at $s$, in the notation of \secref{envsec}, as the smallest value $\kappa_j/M$ such that $T_0$ exits the envelope $E^{\kappa_j}$ at $s$. It can be shown that an equivalent definition is
\begin{equation}\label{pr}\tilde{p}(s)=\frac{\sum_{m=1}^{M-1} \mathbb{I}[R_m \leq R^*_0(s)]  +1}{M};\end{equation}
and that, as we would expect, %clear from \eqref{pp} that $p_+=\min_{s\in\cS}\tilde{p}(s)$, and this fact, together with the validity of the $p$-value \eqref{pp}, imply that
  the  adjusted $p$-values $\tilde{p}(s)$ control the FWER.%\footnote{Cite a reference}

The adjusted $p$-value \eqref{pr} is not really new. The \texttt{fda} package \cite{ramsay2009} for R \citep{R} offers permutation $t$- and $F$-tests for settings (i) and (ii), respectively, of the Introduction \citep[and similar permutation $F$-tests are described by][]{reiss2010}.
  These tests yield pointwise adjusted $p$-values that are related to \eqref{pr}, but there are two differences. First, in the terminology of \cite{ge2003}, the \texttt{fda} package offers \emph{max T} adjusted $p$-values, whereas \eqref{pr} is more akin to \emph{min P} adjusted $p$-values, which are more appropriate when one cannot assume the  null distribution of $T(s)$ to be identical across $s$. Second,  \cite{ramsay2009} adopt a different permutation $p$-value convention in which the numerator and denominator are reduced by 1, leading to the zero $p$-value problem criticized by \cite{phipson2010}.

\section{More powerful $p$-value adjustments}\label{morepower}
We describe next two alternative adjusted $p$-values that are bounded above by \eqref{pr} and thus offer greater power.

\subsection{Step-down adjustment}\label{stepsub}
In the language of multiple testing, the adjusted $p$-values \eqref{pr} are of \emph{single-step} type, suggesting that an analogous \emph{step-down} procedure \citep{westfall1993,ge2003,romano2016} would be more powerful. 
 Define
$S_i=\{s\in\cG: R_0^*(s)\geq i\}$ for $i=1,2,\ldots$, and $R_{m;U}=\min_{s\in U}R_m^*(s)$ for $m\in\{0,\ldots,M-1\}$ and $U\subset \cG$. We can then define the step-down adjusted $p$-value at $s$  as
\begin{equation}\label{pss}
\tilde{p}^{\text{stepdown}}(s)=\max_{i\in\{1,\ldots,R^*_0(s)\}}\frac{\sum_{m=1}^{M-1} \mathbb{I}(R_{m;S_i} \leq i)  +1}{M}. %\mbox{ for }s\mbox{ such that }R^*_0(s)=k
\end{equation}
 This expression is readily shown to be less than or equal to  $\tilde{p}(s)$ in \eqref{pr}. Thus the step-down adjusted $p$-values offer greater power than their single-step counterparts, but they can be shown to retain control of the FWER.

\subsection{Extreme rank length adjustment}\label{erlsub}
\begin{figure}[b]
\centering
\includegraphics[width=\textwidth]{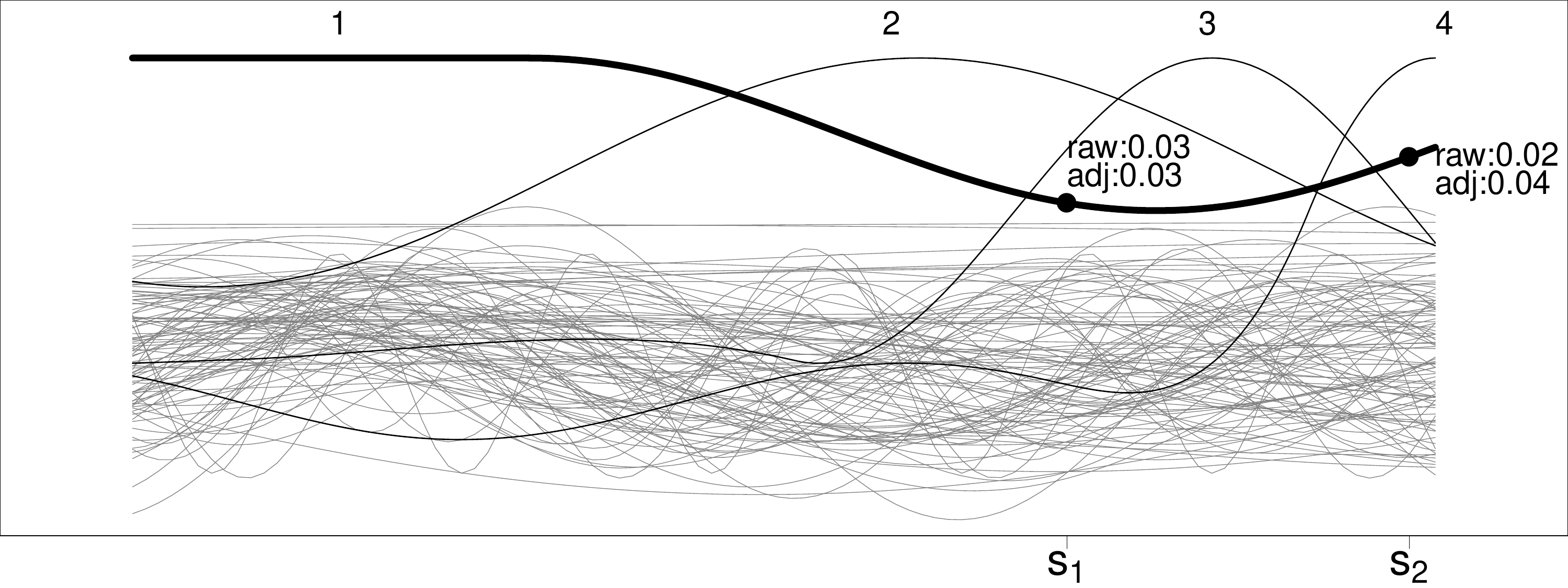}
\caption{An illustration of one-sided (higher = more extreme) ERL depths, and associated pointwise adjusted $p$-values. Here $M=100$ and the numerals 1--4 denote ERL depths for the four curves with $R_m=1$; the thickest curve represents the real data, so that $R_0^{\text{ERL}}=1$. The raw $p$-values \eqref{rawp} %of the two points indicated
  satisfy $p(s_1)>p(s_2)$, but ERL adjustment reverses the order, i.e., $\tilde{p}^{\text{ERL}}(s_1)<\tilde{p}^{\text{ERL}}(s_2)$.}
\label{erlfig}        
\end{figure}

The min-rank depth $R_m$ of \secref{envsec} tends to be strongly affected by ties. In particular, typically $\kappa_1>1$ of the $M$ functions attain rank 1 at some point and thus have $R_m=1$, with the result that $\kappa_1/M$ is the smallest attainable value of either $p_+$ or $\tilde{p}(s)$. An alternative functional depth, the \emph{extreme rank length} (ERL), largely eliminates ties and thus leads to a more powerful variant of the GET. A formal definition of ERL appears in \cite{myllymaki2017}, but the basic idea is to break the tie among curves with the same min-rank depth $R_m$ by ordering from longest to shortest extent of the region over which that minimum rank is attained. For example, four curves  in Fig.~\ref{erlfig} attain pointwise rank~1 (from the top) somewhere in the domain and thus all have $R_m=1$; the ERL depths $R_m^{\text{ERL}}=$1-4, indicated in the figure, are based on the widths of these curves' regions  of attaining rank~1. %\cite{mrkvicka2018} show that the ERL rank leads to definitions of $p$-values and  envelopes such that, as for the minimum rank $R_m$, $T_0$ attains a  given $p$-value threshold if and only if it exits a corresponding envelope.

An ERL envelope $E^{\kappa_j;\text{ERL}}$ \citep{mrkvicka2018} can be defined as in \secref{envsec}, but in terms of $R_m^{\text{ERL}}$ rather than $R_m$. %(Note that if the ERL ranking has no ties then $\kappa_j=j$ for each $j$.)
  We can then proceed as in \secref{adjsec}, and define $\tilde{p}^{\text{ERL}}(s)$, the ERL-adjusted $p$-value at $s$, as $\kappa_j/M$ for the smallest $\kappa_j$ such that $T_0(s)$ lies outside $E^{\kappa_j;\text{ERL}}$. This adjusted $p$-value is bounded above by \eqref{pr}, and hence offers improved power. However,  unlike most $p$-value adjustments, the ERL adjustment is not order-preserving, in the sense that $p(s_1)>p(s_2)$ does not guarantee that $\tilde{p}^{\text{ERL}}(s_1)\geq\tilde{p}^{\text{ERL}}(s_2)$. An counterexample, that is, a pair of points $s_1,s_2$ for which $p(s_1)>p(s_2)$ but $\tilde{p}^{\text{ERL}}(s_1)<\tilde{p}^{\text{ERL}}(s_2)$, appears in Fig.~\ref{erlfig}. Some might argue that this non-order-preserving behavior vitiates the use of ERL-adjusted $p$-values altogether.

\section{Application: Age-varying sex difference in cortical thickness}\label{appsec}

\begin{figure}[b]
\centering
\includegraphics[width=\textwidth]{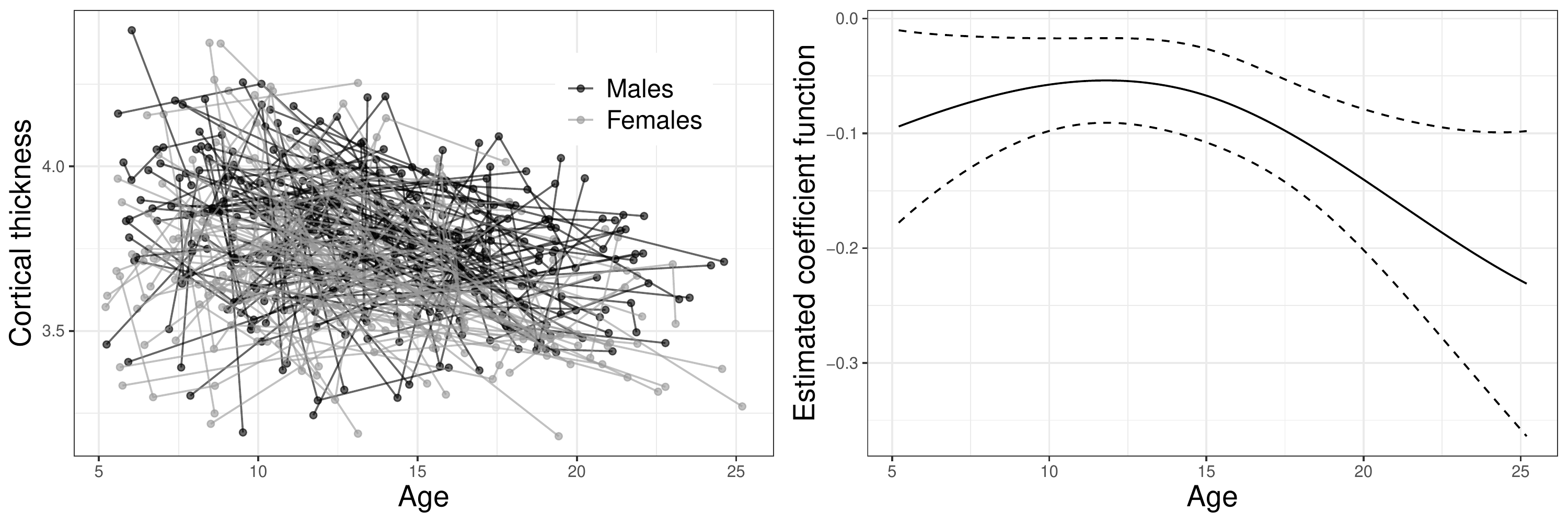}
\caption{Left: Cortical thickness in the right superior temporal gyrus for the NIMH sample. Right: Coefficient function estimate $\hat{\beta}_1(s)$ representing sex effect (female minus male), along with approximate pointwise 95\% confidence interval.}
\label{fig67}        
\end{figure}

We consider cortical thickness (CT) measurements from a longitudinal magnetic resonance imaging study at the US National Institute of Mental Health, which were previously analyzed by \cite{reiss2018}. Specifically, we examine CT in the right superior temporal gyrus in 
 131 males with a total of 355 observations, and 114 females with 300 observations, over the age range from 5--25 years (displayed in the left panel of Fig.~\ref{fig67}). Viewing the observations as sparse functional data, we fit the model
 $y_i(s)=\beta_0(s)+\tau_i\beta_1(s)+\varepsilon_i(s)$, in which $y_i(s)$ is the $i$th participant's CT at age $s$; $\tau_i=0,1$ if this participant is male or female, respectively; and $\varepsilon_i(s)$ denotes error. We focus on testing whether the age-varying sex effect $\beta_1(s)$  (female minus male) equals zero; see the right panel of Fig.~\ref{fig67} for an estimate of this coefficient function, along with pointwise 95\% confidence intervals. 

 \begin{figure}[b]
\centering
\includegraphics[width=\textwidth]{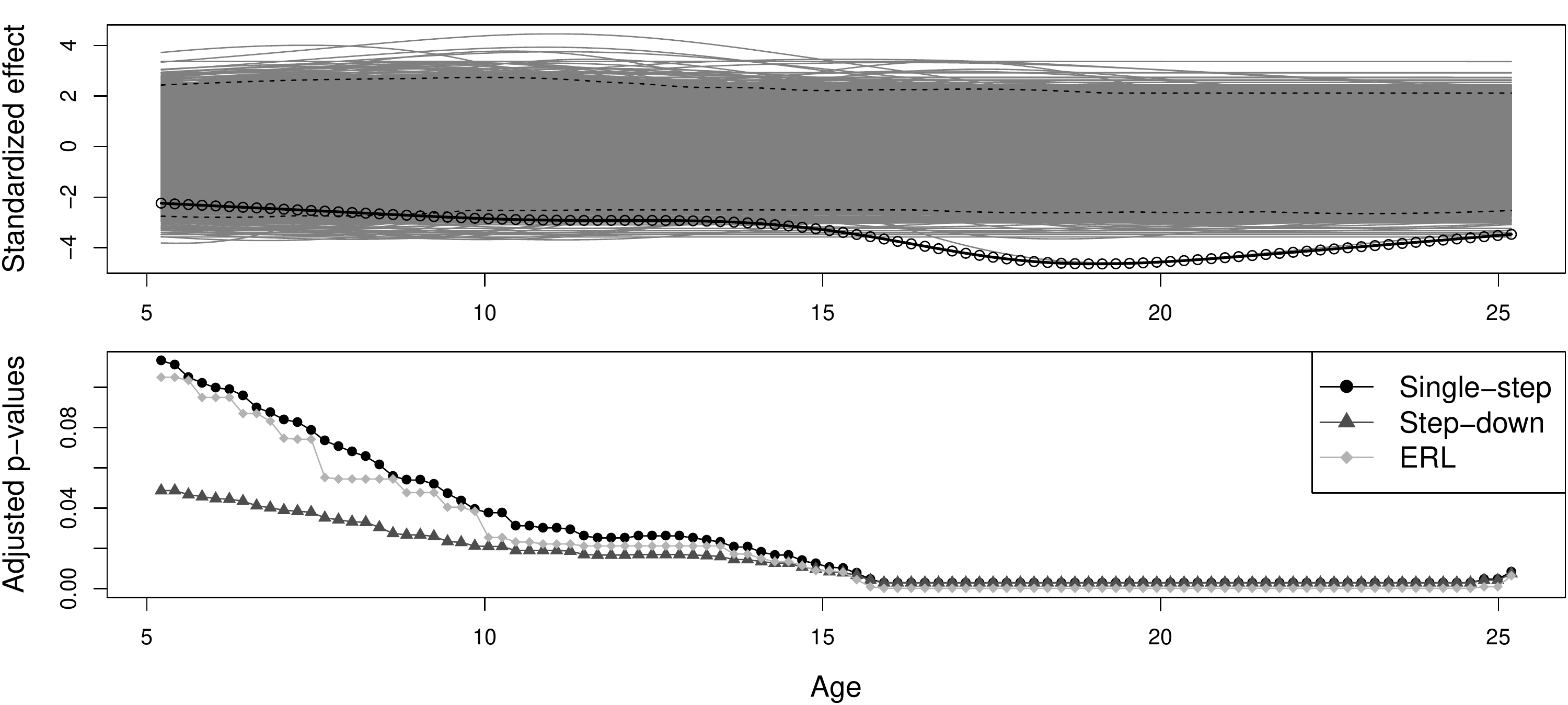}
\caption{Above:  Standardized coefficient functions $\hat{\beta}_1(s)/\mbox{ } \widehat{\textsc{se}}\mbox{ }[\hat{\beta}_1(s)]$ for the real data (black curve and circles) and for 3999 permuted data sets (grey curves), adapted from the R package GET \citep{myllymaki2017}. Dashed lines indicate envelope for testing at the 5\% level. Below: Pointwise adjusted $p$-values $\tilde{p}(s)$ (single-step), $\tilde{p}^{\text{stepdown}}(s)$ and $\tilde{p}^{\text{ERL}}(s)$.}
\label{p67}        
\end{figure}

 The model was fitted by the \texttt{pffr} function \citep{ivanescu2015}, part of the R package \texttt{refund}  \citep{refund}, with both the real data and $M-1=3999$  data sets with the sex labels permuted. The upper panel of Fig.~\ref{p67}  displays standardized coefficient functions $\hat{\beta}_1(s)/\mbox{ } \widehat{\textsc{se}}\mbox{ }[\hat{\beta}_1(s)]$ for the real and permuted data sets, along with a two-sided envelope for testing at the 5\% level.  The GET $p$-value \eqref{pp} based on min-rank depth is $p_+=.003$; if we instead use the ERL depth, the GET $p$-value falls to .00025 ($=1/M$). But to quantify the evidence of a sex effect in an age-specific manner, we require pointwise $p$-values.
 
 %Since the test of interest is two-sided, the pointwise ranks $R_m^*(s)$ determining the envelopes are defined as either rank from the top or from the bottom, whichever is smaller.
  The lower panel of Fig.~\ref{p67}  shows the pointwise adjusted $p$-values $\tilde{p}(s)$ \eqref{pr}, along with the step-down and ERL-based adjusted $p$-values of \secref{morepower}, for an evenly spaced grid of 100 ages. Judging from the values of $\tilde{p}(s)$, there is only weak evidence of a CT difference between girls and boys up to age~9. The step-down $p$-values in this age range, on the other hand, are markedly lower and consistently below the conventional .05 level. The ERL-adjusted $p$-values are closer to $\tilde{p}(s)$ in this lower age range but, somewhat less visibly, are the lowest of the three $p$-values for age 16 and higher. Thus neither one of the two adjustments of \secref{morepower} consistently dominates the other.
  
 It must be acknowledged that the right superior temporal gyrus was specifically selected for the purpose of illustrating differences that may arise among the $p$-value adjustments. Comparable analyses for most other brain regions would have yielded less prominent differences.

\section{Discussion}

Expression \eqref{pr} defines distribution-free pointwise adjusted $p$-values with respect to the global envelope test of \cite{myllymaki2017}. A pointwise $p$-value approach such as this, which is agnostic with respect to the distribution of $T(s)$, is particularly valuable in analyses that go beyond  pointwise $t$- or $F$-tests. For example, we are currently developing flexible pointwise tests for group differences in a measure of interest, based on estimating each group's density at each $s$, and then referring the distance between group-specific densities to a  permutation distribution for each $s$; this distribution has no known analytic form under the null hypothesis.

The step-down and ERL-based adjusted $p$-values of \secref{morepower} offer more powerful alternatives to \eqref{pr}, but some might question the suitability of the ERL adjustment since it is not order-preserving in general. The cortical thickness analysis of \secref{appsec} illustrates the power gains that the step-down and ERL adjustments may provide in some applications. Simulation studies will further elucidate the relative performance of alternative $p$-value adjustments in FDA settings.

\begin{acknowledgement}
This work was supported by Israel Science Foundation grant 1777/16. We thank Aaron Alexander-Bloch, Jay Giedd and Armin Raznahan for providing the cortical thickness data, and for advice on processing these data.
\end{acknowledgement}

\bibliographystyle{chicago}
\bibliography{step}

\begin{thebibliography}{}

\bibitem[\protect\citeauthoryear{Baddeley, Diggle, Hardegen, Lawrence, Milne,
  and Nair}{Baddeley et~al.}{2014}]{baddeley2014}
Baddeley, A., P.~J. Diggle, A.~Hardegen, T.~Lawrence, R.~K. Milne, and G.~Nair
  (2014).
\newblock On tests of spatial pattern based on simulation envelopes.
\newblock {\em Ecological Monographs\/}~{\em 84\/}(3), 477--489.

\bibitem[\protect\citeauthoryear{Cox and Lee}{Cox and Lee}{2008}]{cox2008}
Cox, D.~D. and J.~S. Lee (2008).
\newblock {Pointwise testing with functional data using the Westfall--Young
  randomization method}.
\newblock {\em Biometrika\/}~{\em 95\/}(3), 621--634.

\bibitem[\protect\citeauthoryear{Davison and Hinkley}{Davison and
  Hinkley}{1997}]{davison1997}
Davison, A.~C. and D.~V. Hinkley (1997).
\newblock {\em Bootstrap Methods and Their Application}.
\newblock Cambridge University Press.

\bibitem[\protect\citeauthoryear{Ge, Dudoit, and Speed}{Ge
  et~al.}{2003}]{ge2003}
Ge, Y., S.~Dudoit, and T.~P. Speed (2003).
\newblock Resampling-based multiple testing for microarray data analysis (with
  discussion).
\newblock {\em TEST\/}~{\em 12\/}(1), 1--77.

\bibitem[\protect\citeauthoryear{Goldsmith, Scheipl, Huang, Wrobel, Gellar,
  Harezlak, McLean, Swihart, Xiao, Crainiceanu, and Reiss}{Goldsmith
  et~al.}{2018}]{refund}
Goldsmith, J., F.~Scheipl, L.~Huang, J.~Wrobel, J.~Gellar, J.~Harezlak, M.~W.
  McLean, B.~Swihart, L.~Xiao, C.~Crainiceanu, and P.~T. Reiss (2018).
\newblock {\em refund: Regression with Functional Data}.
\newblock R package version 0.1-17.

\bibitem[\protect\citeauthoryear{Ivanescu, Staicu, Scheipl, and
  Greven}{Ivanescu et~al.}{2015}]{ivanescu2015}
Ivanescu, A.~E., A.-M. Staicu, F.~Scheipl, and S.~Greven (2015).
\newblock Penalized function-on-function regression.
\newblock {\em Computational Statistics\/}~{\em 30\/}(2), 539--568.

\bibitem[\protect\citeauthoryear{L{\'o}pez-Pintado and Romo}{L{\'o}pez-Pintado
  and Romo}{2009}]{lopez2009}
L{\'o}pez-Pintado, S. and J.~Romo (2009).
\newblock On the concept of depth for functional data.
\newblock {\em Journal of the American Statistical Association\/}~{\em 104},
  718--734.

\bibitem[\protect\citeauthoryear{Mrkvi{\v{c}}ka, Myllym{\"a}ki, Jilek, and
  Hahn}{Mrkvi{\v{c}}ka et~al.}{2018}]{mrkvicka2018}
Mrkvi{\v{c}}ka, T., M.~Myllym{\"a}ki, M.~Jilek, and U.~Hahn (2018).
\newblock {A one-way ANOVA test for functional data with graphical
  interpretation}.
\newblock arXiv preprint arXiv:1612.03608.

\bibitem[\protect\citeauthoryear{Myllym{\"a}ki, Mrkvi{\v{c}}ka, Grabarnik,
  Seijo, and Hahn}{Myllym{\"a}ki et~al.}{2017}]{myllymaki2017}
Myllym{\"a}ki, M., T.~Mrkvi{\v{c}}ka, P.~Grabarnik, H.~Seijo, and U.~Hahn
  (2017).
\newblock Global envelope tests for spatial processes.
\newblock {\em Journal of the Royal Statistical Society: Series B\/}~{\em
  79\/}(2), 381--404.

\bibitem[\protect\citeauthoryear{Phipson and Smyth}{Phipson and
  Smyth}{2010}]{phipson2010}
Phipson, B. and G.~K. Smyth (2010).
\newblock {Permutation $p$-values should never be zero: calculating exact
  $p$-values when permutations are randomly drawn}.
\newblock \emph{Statistical Applications in Genetics and Molecular Biology},
  9(1), article 39.

\bibitem[\protect\citeauthoryear{{R Core Team}}{{R Core Team}}{2019}]{R}
{R Core Team} (2019).
\newblock {\em R: A Language and Environment for Statistical Computing}.
\newblock Vienna, Austria: R Foundation for Statistical Computing.

\bibitem[\protect\citeauthoryear{Ramsay, Hooker, and Graves}{Ramsay
  et~al.}{2009}]{ramsay2009}
Ramsay, J.~O., G.~Hooker, and S.~Graves (2009).
\newblock {\em {Functional Data Analysis with R and MATLAB}}.
\newblock New York: Springer.

\bibitem[\protect\citeauthoryear{Reiss}{Reiss}{2018}]{reiss2018}
Reiss, P.~T. (2018).
\newblock Cross-sectional versus longitudinal designs for function estimation,
  with an application to cerebral cortex development.
\newblock {\em Statistics in Medicine\/}~{\em 37\/}(11), 1895--1909.

\bibitem[\protect\citeauthoryear{Reiss, Huang, and Mennes}{Reiss
  et~al.}{2010}]{reiss2010}
Reiss, P.~T., L.~Huang, and M.~Mennes (2010).
\newblock Fast function-on-scalar regression with penalized basis expansions.
\newblock \emph{International Journal of Biostatistics}, 6(1), article 28.

\bibitem[\protect\citeauthoryear{Ripley}{Ripley}{1977}]{ripley1977}
Ripley, B.~D. (1977).
\newblock Modelling spatial patterns.
\newblock {\em Journal of the Royal Statistical Society: Series B\/}~{\em
  39\/}(2), 172--192.

\bibitem[\protect\citeauthoryear{Romano and Wolf}{Romano and
  Wolf}{2016}]{romano2016}
Romano, J.~P. and M.~Wolf (2016).
\newblock Efficient computation of adjusted $p$-values for resampling-based
  stepdown multiple testing.
\newblock {\em Statistics \& Probability Letters\/}~{\em 113}, 38--40.

\bibitem[\protect\citeauthoryear{Westfall and Young}{Westfall and
  Young}{1993}]{westfall1993}
Westfall, P.~H. and S.~S. Young (1993).
\newblock {\em Resampling-Based Multiple Testing: Examples and Methods for
  $P$-Value Adjustment}.
\newblock New York: John Wiley \& Sons.

\bibitem[\protect\citeauthoryear{Wright}{Wright}{1992}]{wright1992}
Wright, S.~P. (1992).
\newblock Adjusted $p$-values for simultaneous inference.
\newblock {\em Biometrics\/}~{\em 48\/}(4), 1005--1013.

\end{thebibliography}

\end{document}